# WE model: A Machine Learning Model Based on Data-Driven Movie Derivatives Market Prediction

Yaoyao Ding[1], Chenghao Wu[2], Yuntao Zou[3], Peng Zhou[4]

1. Yaoyao Ding , Macau University of Science and Technology, Macau, 999078, China ,2109853ZAM30002@student.must.edu.mo
2. Chenghao Wu , Macau University of Science and Technology, Macau, 999078, China ,wuchenhao78@gmail.com
3. Yuntao Zou , Huazhong University of Science and Technology, , Wuhan, 430074, China ,zouyuntao@hust.edu.cn
4. Peng Zhou , Huazhong University of Science and Technology, , Wuhan, 430074, China ,peng_zhou@hust.edu.cn

# Abstract

The mature development and the extension of the industry chain make the income structure of the film industry. The income of the traditional film industry depends on the box office and also includes movie merchandising, advertisement, home entertainment, book sales etc. Movie merchandising can even become more profitable than the box office. Therefore, market analysis and forecasting methods for multi-feature merchandising of multi-type films are particularly important. Traditional market research is time-consuming and labour-intensive, and its practical value is restricted. Due to the limited research method, more effective predictive analysis technology needs to be formed. With the rapid development of machine learning and big data, a large number of machine learning algorithms for predictive regression and classification recognition have been proposed and widely used in product design and industry analysis. This paper proposes a high-precision movie merchandising prediction model based on machine learning technology: WE model. This model integrates three machine learning algorithms to accurately predict the movie merchandising market. The WE model learns the relationship between the movie merchandising market and movie features by analyzing the main feature information of movies. After testing, the accuracy rate of prediction and evaluation in the merchandising market reaches 72.5%, and it has achieved a strong market control effect.

# 1.Introduction

## 1.1. Research Content

This research pioneered a machine learning algorithm model for movie merchandising market prediction: WE model. Through the extensive collection of movie data, a set of datasets for the study of movie merchandising is created, and then the dataset is marked. A large number of industry experts are invited to use the Delphi method to evaluate and score movie merchandising of the sample data, giving the movie dataset true mathematical research value. Finally, the WE model is used to train and predict the algorithm model on the labelled movie dataset.

Introducing new research methods and forecasting models for movie merchandising, on the one hand, can help production companies to do a good job in the development planning of merchandising in the early stage. On the other hand, film development merchandising helps manufacturers and retailers best invest in risk management to select the films with the most potential to develop merchandising instead of blindly entering the market. The vigorous development of the film industry has accelerated the linkage between film and other industries, making the film begin to have a "chemical reaction" with other brands and participate in a broader cultural consumption structure. Movie merchandising is no longer a subsidiary product of movies but an indispensable part of the movie industry and has profoundly changed the marketing methods and revenue structure of the movie industry.

Although movie merchandising can add a steady stream of revenue to movies, not all movie merchandising can be recognized by the market. As time passes, the trend of "oligopoly" has gradually become significant: in addition to some popular movie characters being sought after by the market, other films have not been highly recognized by the market. It can be said that the market access threshold is very high, but the success rate is very low. The main reasons are based on three points:

1. The development of movie merchandising involves a more complex market system than movie projects. Because the subjects involved are enormous and complex, the products of each category and the consumer groups targeted are vastly different, which increases the difficulty of evaluation.

2. In the era of product proliferation, consumer preferences vary in real-time, and successful movie merchandising cases cannot necessarily be copied, which increases the difficulty of development;

3. Film development merchandising may involve a series of factors, such as the development of film characters, elements, and content and the corresponding product design, all of which are difficult to quantify and evaluate.

The exploration and study of movie merchandising has a long history. Disney is a pioneer in this field. As early as when Disney was a small animation company, founder Walter Disney had already started the *Mickey Mouse* merchandising licensing activities to obtain the "long tail income" of the movie.

The major studios initially did not attach importance to the market value of merchandising. *Star Wars* (1977) even gave away the licensing rights to merchandisingrs who would commit themselves to advertising campaigns for the products. While it may have attracted a number of movie audiences for *Stars Wars*, missed the potential licensing income that might have been made on the movie characters. "As it turned out, more than $4 billion worth of *Star Wars*-related merchandising was sold, which, at a 6 per cent royalty on the wholesale price, would have produced $120 million in licensing fees." (Edward, 2005) [1] This "merchandising tie-in" strategy (Wyatt, 1994)[2] helps Hollywood studios see huge business opportunities. In the subsequent film marketing strategy, a comprehensive marketing system with film copyright as the core and driving the common development of other industries will be built.

This strategy has changed the revenue structure of the film and promoted the great development of the licensing market. According to Licensing International's annual survey, the sixth consecutive year of growth in 2019 sets a solid foundation for the industry ahead of the pandemic. The global sales of licensed merchandising and services grew to $292.8 billion in 2019, a 4.5% increase over the $280.3 billion generated in 2018.

This also shows that in recent years, licensed merchandising and services have ushered in historical opportunities, and their influences on the film industry have gradually expanded, especially in the special context of the pandemic, the global licensing market has a strong and profitable business model and distanced itself from other industries.

Among them, the entertainment/character sector remained the leading market share category by far, accounting for $128.3 billion, or 43.8% of the total global licensing market. Apparel (15.1%), toys (12.2%) and fashion accessories (11.9%) continued to lead in the breakdown by product category (Samantha, 2020) [3]. These categories provide an important reference for merchandising research in this paper.

According to the "Top Global Licensors 2021" report [4], not surprisingly, seven of the top ten licensors are entertainment companies, occupying an absolute advantage in the licensing market. Among them, The Walt Disney Company topped the list with $54 billion in sales, WarnerMedia, Hasbro, NBCUniversal/Universal Brand Development, ViacomCBS and so on have a sizable market share. These companies emerged from the chaotic scenes of the pandemic and finally managed to "weather the storm". Among them, the traditional toy company Hasbro, as well as its competitors Mattel, LEGO and MGA Entertainment, have rebranded themselves as entertainment companies over the past decade or so. Hasbro operates a series of movie-related toys such as Transformers, Barbie, Special Forces, etc., which not only creates long-term brand value but also manages merchandising as an important part of the cultural consumption structure.

## 1.2. Research Status

Research in academia lags far behind the industry. In the decades after the birth of Mickey Mouse, few scholars have paid attention to movie merchandising. Until after the 1980s, move merchandising stated to be discussed in the film and media world. Research in these two fields often links to cultural and economic trends, exploring a range of issues such as consumer culture, brand values and capitalism.

In the context of consumer culture in the 20th century, movies are naturally associated with department stores. Jane Gaines (1989) [5] pointed out that the display window is like a movie screen, not only a place for showing the movie merchandising, but also a bridge between the soft screen aesthetic and hard consumer; Charles Eckert (1978) [6] pointed out that Hollywood has given a unique tendency to consumerism, and has a lot of associations with fashion, cosmetics, department stores, etc. It is said that movie theatres and display windows have become consumer fantasy places for modern leisure and entertainment. Nowadays, the addition of the Internet and digital culture has further expanded the space for cultural consumption of movie merchandising.

Some scholars have also noticed that the value of movie merchandising to the entertainment industry has begun to affect how movies tell stories and audience engagement strategies, and fans research has naturally become a core concern in this field. Matt Hills (2003) [7] related the concept of fan-consumer to the concept of fan-producer. Jonathan Gray (2010) [8] pointed out that movie merchandising can not only stimulate audience expectations but also make audiences have film imagination and cultural resonance. As gender is often the case with discussions of merchandising, female celebrities and feminized products are the core of this work. Derek Johnson (2014)[9] took Star Wars as an example to dissect the industrial logic of media franchises, post-feminist culture, and gender and sexuality in social media.

Although movie merchandising has long been an important topic in film and media research, these studies are limited to the field of qualitative research, mainly using case analysis, questionnaire survey, historical context and other research methods. It has played a certain role in helping us understand the concept, development process, marketing model, and consumer culture of movie merchandising [10], but it has not solved the problem of "how ". In a sense, movie merchandising is the product of the development of the film industry, involving a series of more complex business models, trade, finance, investment, production, statistics and other issues. Therefore, the introduction of quantitative research methods has become particularly important.

In this paper, a set of movie datasets is constructed for the development of combining movie merchandising. Due to the limited samples included, there are certain obstacles to the research of artificial intelligence algorithms. In order to solve the problem of small sample data, this paper adopts a large number of classic machine learning algorithms, which have achieved good results in the value prediction of movie merchandising.

# 2. Machine Learning Models

To conduct a quantitative analysis of the movie merchandising, it is necessary to establish a mathematical model. Mathematical modeling is a mathematical thinking method. It uses mathematical language and methods to approximate and solve practical problems through abstraction and simplification, which is a powerful mathematical tool.

The main contribution of this paper is that we have initially created a research method: using movie datasets to build a data model and use it for artificial intelligence algorithm analysis and prediction of movie merchandising.

## 2.1. Movie dataset

A mathematical model is generally a mathematical simplification of the real thing. It often exists in an abstract form that is close to the real thing in a sense, but it is essentially different from the real thing. [11] In order to make the description more scientific, logical, objective and repeatable, we need to collect the data of related movie merchandising as comprehensively as possible and build a set of movie-related data sets for establishing the movie data model.

First of all, the machine learning model proposed in this study is a data-driven model, and the amount of data will have a great impact on the prediction effect of this model. Therefore, when constructing movie datasets as training, validation, and testing sets, large-sample datasets have better support for the robustness of the model and can avoid overfitting. Movie merchandising licensing activities did not develop rapidly until the late 20th and early 21st centuries, dating back to 1977 when Twentieth Century–Fox released *Star Wars*. Therefore, the collection time of movie information data should start from the 1970s to the present. At the same time, movies with a certain influence are required for the selection of movie data samples. We cannot deny that an influential film does not necessarily have the exploitability value of movie merchandising, but a film must have a high influence to get profits in the licensing market. There are many various dimensions to judge the influence of a film, such as the box office amount, number of moviegoers, release area, audience popularity, etc. This paper will select movies with high influence based on certain measurement indicators when constructing a movie dataset. Data is collected and labelled to ensure model validity.

Secondly, the information of movie merchandising needs to have a high enough dimension. Although high-dimensional data can greatly improve the difficulty of training, it can reflect more comprehensive movie merchandising feature information. The success of movie merchandising is a multi-dimensional and multi-factor problem, and it is difficult to directly express through certain data. When collecting movie data, we did not know how the data factor plays an important role in the development value of movie merchandising at the beginning. For example, the film *Inception* is a popular movie, in which there is also a rotating gyroscope, a prop impressed by the audience. However, this film did not create enough "merchandising value" and carry out a large number of merchandising development projects. So it is necessary to collect relevant data from the film as comprehensively as possible.

## 2.2. Data preprocessing method

The original movie dataset obtained through data collection and integration is incomplete, and it is necessary to preprocess the data to obtain a labelled dataset that can be used for machine learning model training and testing [12]. The essence of data preprocessing is to transform rough, low-quality data sets, and a series of data cleaning operations into smooth, high-quality data sets. An interesting analogy for data preprocessing: serial data engineering of preprocessing is a process that transforms pure mathematicians, i.e. raw data sets, into physicists, biologists, economists, and even film industry scientists so that the new data set can participate in the training and adjustment of the model in various prediction regression, classification and recognition problems.

A more vivid metaphor for data preprocessing is that the original data set is a pure mathematician, and the data information contained in it is high-dimensional, multi-category, and often high-level black-box problems. In the research of this paper, the essence of the data preprocessing method is similar to the training of the mathematician through the film industry experts, teaching him the knowledge of the film industry and letting him know the meaning and weight of various factors in the film data set, that is, the original film data. The sets are combined, cropped, and arranged in an orderly manner, and the movie data samples are labelled, which can be directly used for the training of machine learning models and the adjustment of hyperparameters to improve the efficiency of data utilization.

## 2.3. Labelled dataset construction based on the Delphi method

In the supervised learning task, to better train the model, it is necessary to manually label the samples, and the labelled data set is the basis for the machine learning model to update the parameters and improve the prediction accuracy. Since the accuracy of labelling will directly affect the training effect of the entire model, how to label properly becomes the primary problem to be solved. This paper uses the Delphi method to label the dataset.

The success of movie merchandising is a multi-dimensional and multi-factor issue, which needs to be represented by a variety of data, and evaluated and marked according to the movie sample data combined with the actual value of the movie merchandising. Therefore, this paper adopts the Delphi method, an effective and reliable method for collecting expert opinions, which was developed by the American RAND Corporation and Douglas Corporation in the 1950s. The core of this method is to consult the experts separately, and there is no communication between the experts during the evaluation; this can integrate the opinions of each expert without allowing the experts to be interfered with by others; the evaluation results generated in the previous round are fed back to all experts, and then repeat the next round of evaluation; through several rounds of feedback, the various experts' opinions are gradually reached. The flowchart of the Delphi method is shown in Figure 2-1.

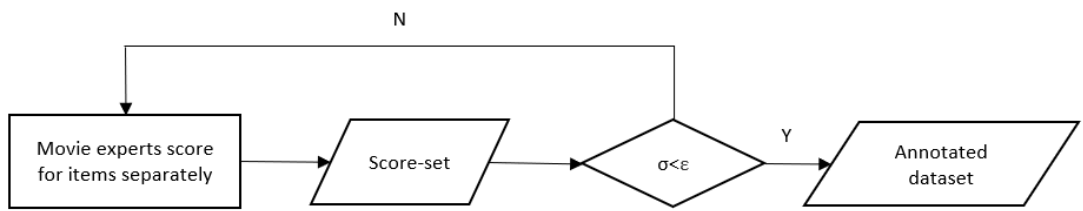

Figure 2-1 Delphi method flow

It can be seen from the flow chart of the Delphi method that the process of expert scoring requires multiple iterations, where σ is the variance of all expert scores in each round, and ε is the self-set variance deviation value. When the scoring variance σ of all experts is less than the set. When the value of ε is fixed, we think that the expert opinions tend to be consistent and get the final labelled dataset.

The benefits of using the Delphi method for the production of labelled datasets are:
1. Each expert evaluates independently to avoid authoritative experts influencing others' evaluations
2. Reaching a consensus through multiple rounds of evaluation can prevent some introverted experts from colliding with powerful experts and not expressing their opinions.
3. All evaluation results are presented anonymously, to prevent experts from not correcting their own mistakes for the sake of face.

Combined with the actual application scenarios of this paper, the pseudocode of the Delphi method is as follows:

algorithm：Delphi

Input: Expert-set $\{x_1, x_2, x_3, ......x_i\}$  x: movie expert

Input: Score-set $\{f_j(x_1,k), f_j(x_2,k), f_j(x_3,k)...... f_j(x_i,k)\}$   $f_j(x_i,k)$: scores of movie experts for movie sample k at the round j.

$$\sigma = \sqrt{\frac{1}{n}\sum_{i=1}^{n}(f_j(x_i, k) - M)^2} \qquad M = \frac{1}{n}\sum_{i=1}^{n} f_j(x_i, k)$$

While  $(\sigma < \varepsilon)$  do         // $\varepsilon$: Standard deviation for judgment

movie experts scores for every movie sample, as $f_j(x_i, k)$;

summary the scores;

calculate the value of $\sigma$ ;

End While

## 2.4. Movie merchandising prediction model algorithm

### 2.4.1. Difficulty Analysis of Algorithm Model

After completing the production of the labelled data set by introducing the Delphi method, the problem studied in this paper will be determined as a supervised learning problem. Based on the analysis of the characteristics of the labelled data set after data preprocessing, this paper first summarizes the difficulties of this problem combined with the data situation.

Firstly, the distribution map of the value score of movie merchandising shown in Figure 2-2 is given. It can be seen from the figure that there is too much low-quality evaluation data, and there is a large number of sample data with a value score of 0. At the same time, the distribution of merchandising value scores of movie samples in the movie dataset is especially extreme, and there are cases where scores are missing. Of the total of 441 samples, the samples with a score of 0 account for about half of the total number of samples. A score of 0 means that the value of movie merchandising is very low. In the problem of predicting the relationship between movie feature information and movie merchandising market conditions, it is impossible to play the role of model adjustment.

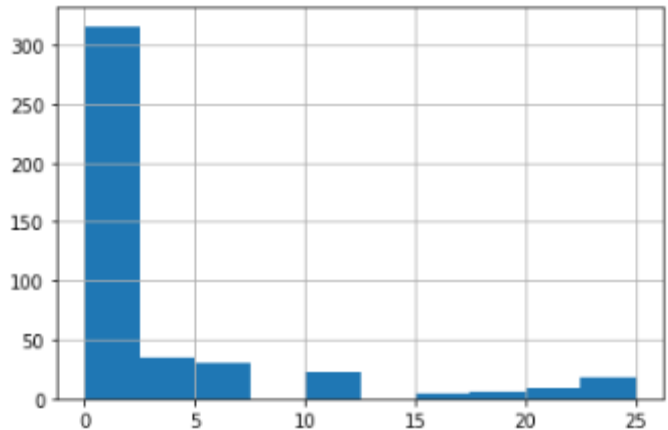

Figure 2-2 Value Score Distribution of Movie Merchandising

The essence of a predictive model is an end-to-end mapping relationship. Data samples with a score of 0 often exist in the form of high-dimensional sparse matrices in the prediction model. When the machine learning model performs gradient backpropagation and other calculations, problems such as gradient disappearance are prone to occur, which reduces the accuracy of the model and affects the effect of artificial intelligence algorithms.

Secondly, although the training of machine learning models needs to be driven by data, it is actually difficult to achieve large data sets in a broad sense. Movie works are limited and cannot support the number of samples in large datasets. We collected a total of 441 movie samples. This number of samples is small for the training of machine learning models and will also affect the training effect of artificial intelligence algorithms.

Quantitative research and machine learning algorithms for movie merchandising in the current research field are still in their infancy, and mature algorithms have not yet been proposed. Existing research mainly focuses on movie box office prediction, rather than the industry analysis of movie merchandising. Although this has a certain reference for us, the methods available for reference are still very limited. [18][19][20][21][22]

To sum up, according to the characteristics of the movie dataset, we select the following four machine learning algorithm models as sub-models of the WE model.

**2.4.2. Linear regression model**

Linear regression is a supervised learning algorithm, that is, there is a training set in the learning, the relationship between the input and the output is known, and the input data has a label, which is used to predict the linear relationship between the dependent variable and the multivariate independent variable. [16][17]

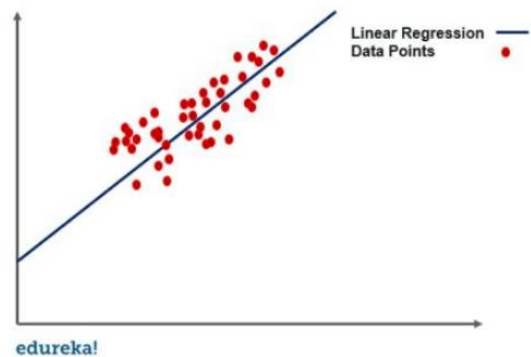

Figure 2-3 Linear regression fitting

We predict the outcome of the dependent variable based on the independent variable, where the independent variable has a linear relationship with the dependent variable, the straight line in the graph is the line of best fit, which can be based on the linear equation given below:

$$Y = b_0 + b_1 x + e \qquad (2-1)$$

The cost function provides the best possible values for b_0 and b_1 in order to make the best fit line for the data points. We obtain the optimal values of b_0 and b_1 by transforming this problem into a minimization problem. In this problem, the error between the actual value and the predicted value is minimized.

$$\min \frac{1}{n} \sum_{i=1}^{n} (pred_i - y_i)^2 \qquad (2-2)$$

We choose the function above to minimize the error. The error difference is squared and the error for all data points is summed, i.e. the division between the total number of data points. The resulting value then provides the mean squared error for all data points. It is also called MSE (Mean Squared Error) and changes the values of b_0 and

b_1 to stabilize the MSE value at the minimum value.

### 2.4.3. XGBoost Model

XGBoost (Extreme Gradient Boosting) is a supervised learning model, and the model corresponding to XGboost is essentially a bunch of CART trees. Use a bunch of trees to make predictions, and train the next tree on the basis of training one tree to predict the gap between it and the true distribution. Through continuous training of trees used to bridge the gap, a combination of trees is finally used to simulate the real distribution

The big difference between XGBoost and GBDT is the definition of the objective function. The objective function of XGboost includes two parts: loss function and regular term.

$$\mathscr{L}^{(t)} = \sum_{i=1}^{n} l\left(y_i, \hat{y}_i^{(t-1)} + f_t(x_i)\right) + \Omega\left(f_t\right) \qquad (2-3)$$

Additionally: l is the loss function and $\Omega(ft)$ is the regularization term.

The XGBoost model is good at capturing the dependencies between complex data, can obtain effective models from large-scale data sets, and supports a variety of systems and languages in terms of practicability, but its performance on high-latitude sparse feature data sets and small-scale data sets is not very good.

### 2.4.4. LightGBM Model

LightGBM is an implementation of optimizing GBDT (Gradient Boosting Decision Tree), which is an enduring model in machine learning. Its main idea is to use weak classifiers (decision trees) iteratively train to obtain the optimal model.The model has the advantages of good training effect and not easy to overfit

The optimization of LightGBM for XGBoost mainly includes three points: one is the Histogram algorithm, which reduces the number of candidate split points; the other is the GOSS algorithm, a gradient-based unilateral sampling algorithm, which reduces the number of samples; the third is the EFB algorithm, which is mutually exclusive.

Bundling algorithms to reduce the number of features. Therefore, we can think that LightGBM is a combination of four models: XGBoost, Histogram, GOSS, and EFB.

### 2.4.5. LASSO Model

The LASSO model is a regression model, which is a compressed estimate. It obtains a more refined model by constructing a penalty function, so that it compresses some regression coefficients, that is, forcing the sum of the absolute values of the coefficients to be less than a certain fixed value; at the same time, it sets some regression coefficients to zero. Therefore, the advantage of subset shrinkage is preserved, which is a biased estimation for dealing with complex collinear data.

The basic idea of LASSO is to minimize the residual sum of squares under the constraint that the sum of the absolute values of the regression coefficients is less than a constant, so that some regression coefficients strictly equal to 0 can be generated, and an interpretable model can be obtained. Its mathematical expression the formula is as follows:

$$B_{LASSO} = arg_B \min \left\{ \left| Y - \sum_{j=1}^{p} X_j B_j \right| \right\} \qquad (2-4)$$

$$s.t. \sum_{j=1}^{p} |B_j| \leq t$$

Among them, t > 0 is the adjustment parameter, and the overall regression coefficient can be compressed by controlling the adjustment parameter t. The determination of the t value can be estimated using the cross-validation method.

The above four machine learning algorithms are widely used and verified machine learning algorithms. We will use them in the experiments to predict the value of movie merchandising, and compare the prediction accuracy with the WE model proposed in this paper.

The prediction of the movie merchandising market is a brand-new subject. Due to the complexity of the subject itself, we have adopted a variety of technical means. Therefore, we have spent a large amount of introduction in this part. Understanding these technical principles is helpful for the development of the experiment.

# 3. Experiments

Based on the discussed model, experiments are carried out to verify the effect of the model. The experiment is divided into four parts. The first part is to create a movie dataset, the second part is to label the dataset, the third part is data preprocessing, and the fourth part is machine learning algorithm training. Figure 3-1 is the experimental flow chart:

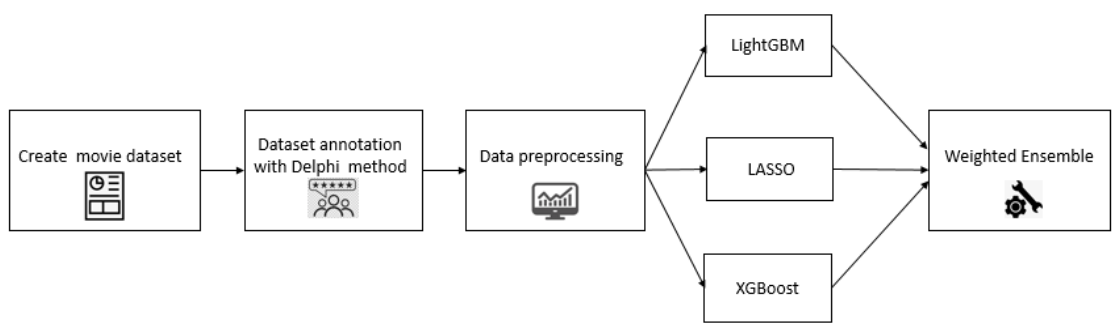

Figure 3-1 Experimental flowchart

As can be seen from the experimental flow chart, after the movie dataset is constructed, experts in the movie industry are invited to rate the value of movie merchandising as a label for machine learning. In the model training stage, three classic machine learning algorithms will be used for training respectively, and finally the obtained model is weighted and integrated to improve the prediction accuracy. This algorithm model is named WE model (WeightedEnsemble model).

## 3.1. Data preprocessing

### 3.1.1. Dataset Generation

This paper collects more than 400 movies of different genres and themes to make a dataset. The data set includes various data as shown in Table 3-1. The movies we select are all influential movies. For movies with great influence, the sales of their merchandising may not be good, but the sales of movie merchandising must be good. Influential films; we also take care to cover as many release years, languages, and regions as possible when selecting films, so that the film sample is broad and representative.

Table 3-1 Movie Information

| Movie Information | | |
|---|---|---|
| Film | Year | Motion Picture Rating(MPAA) |
| Time | IMDB rating | Genres |
| Directors | Writers | Stars |
| Countries of origina | Languages | Filming locations |
| Production companies | Box Office | Script |
| Movie series | How many movie series | |

It can be seen from Table 3-2 that the collected movie sample data basically contains all movie-related information, and comprehensive data helps machine learning achieve good results. In addition, the distribution of movie data samples should be broad and cover more samples.

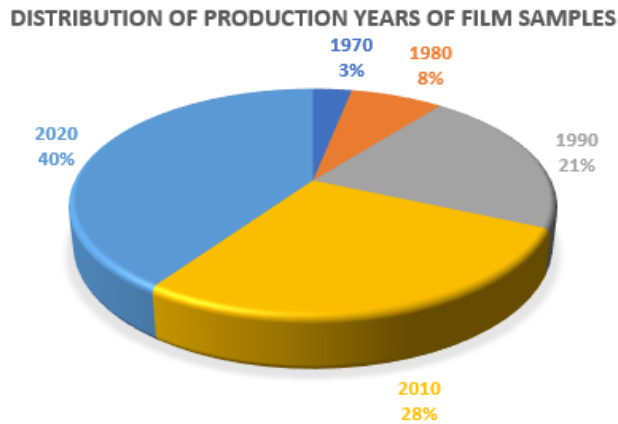

Figure 3-2 Distribution of production years of film samples

It can be seen from Figure 3-1 that the movie sample data covers the period from the 1970s to the 2020s. The more recent movies have more influence, the larger the proportion of the collection is. The movie samples in the 1970s only account for 3%, while The 2020s movie sample is 40%.

### 3.1.2. Dataset Labelling

The supervised learning algorithm we employ requires that each movie sample instance consists of an input object (usually a vector) and the desired output value (also known as a supervised signal), so we need to use the Delphi method for its movie merchandising The situation is evaluated and marked. We invited 20 film industry experts, including directors, producers, distributors, film critics, etc., to rate the sales value of each film in toys, stationery, daily necessities, clothing, bags and other movie merchandising. Table 3-2 lists the metrics for movie sample evaluation. Table 3-2 lists the indicators for evaluating the value of movie Merchandising.

Table 3-2 Summary of Film Sample Evaluation

|  | Movie 1 | | | | | |
|--|--|--|--|--|--|--|
|  | Toys | Stationery | Daily use | Clothes & Accessories | luggage and bags | Total |
| Round 1 | | | | | | |
| Round 2 | | | | | | |
| …… | | | | | | |
| Round n | | | | | | |

Table 3-3 Value Evaluation Form of Movie Merchandising

|  | Round n | | | | | |
|--|--|--|--|--|--|--|
|  | Toys | Stationery | Daily use | Clothes & Accessories | luggage and bags | Total |
| Movie 1 | | | | | | |
| Movie 2 | | | | | | |
| …… | | | | | | |
| Movie n | | | | | | |

After one round of scoring, the scoring results of all experts will be anonymously fed back to each expert, and then the next round of scoring will begin. It can be seen from the flow chart of the Delphi method that the expert scoring needs to be performed multiple times, where σ is the variance of all expert scores in each round, and ε is a value set by yourself. If the consistency requirements for scoring are not strict, the value of ε can be Set to a large number; set ε to a small number if you want the scores of all experts to differ little. When the scoring variance σ of all experts is less than the set value ε, we believe that the opinions of experts tend to be consistent, and the scores obtained at this time can be used as the merchandising value annotation of movie samples.

We set the value of ε to be 2. When the standard deviation σ of the merchandising scores of 20 experts on a movie sample is less than 2, we think that the scores of the experts on the merchandising of this movie sample tend to be consistent, and this score can be used as the movie sample. Annotated value of the sample. Experts must rate all movie samples separately in each round. When the standard deviation of the rating of a movie sample is less than 2, the label value of this movie has been generated, and the next round will not participate in the evaluation.

After 6 rounds of evaluation, the experts' scoring results for all movie samples tend to be consistent, and all samples in the movie dataset complete the evaluation and annotation of movie merchandising.

Table 3-4 lists 3 movie sample data examples, in which experts rated the merchandising market value of the movie *Spider-Man: Homecoming* as 20 points, and it is movie merchandising in the evaluation system with a full score of 25 points. This is a great value movie sample. We observed samples of movies with a high-value evaluation of movie merchandising and found that such movies are usually series, and the total number of series is more than 5, indicating that "How many movie series" and related fields are very important information, below Attention should be paid to the data preprocessing, and detailed information should be supplemented if missing information is found. Similarly, through preliminary analysis, we also found that "Genres" and "MPAA" are also closely related to the value of movie merchandising, which should be paid attention to in the data preprocessing stage.

Table 3-4 Movie Example

| Movie | *Spider-Man: Homecoming* | *The Predator* | *Unbroken* |
|---|---|---|---|
| Year | 2017 | 2018 | 2014 |
| Motion Picture Rating (MPAA) | PG-13 | R | PG-13 |
| Time | 2h13m | 1h47m | 2h17m |
| IMDB rating | 7.4 | 5.3 | 7.2 |
| Genres | Action, Adventure, Sci-Fi | Action, Adventure, Sci-Fi | Action, Biography, Drama |
| Directors | Jon Watts | Shane Black | Angelina Jolie |
| Writers | Jonathan Goldstein, John Francis Daley, Jon Watts | Fred Dekker, Shane Black, Jim Thomas | Joel Coen, Ethan Coen, Richard LaGravenese |
| Stars | Tom Holland, Michael Keaton, Robert Downey Jr. | Boyd Holbrook, Trevante Rhodes, Jacob Tremblay | Jack O'Connell, Miyavi, Domhnall Gleeson |
| Countries of origin | United States | United States, Canada | United States |
| Languages | English, Spanish | English, Spanish | English, Japanese, Italian |
| Filming locations | Berlin, Germany | Vancouver, British Columbia, Canada | Blacktown International Sportspark, Blacktown, Sydney, New South Wales, Australia |
| Production | Columbia Pictures, | Twentieth Century | 3 Arts Entertainment, |

| companies | LStar Capital, Marvel Studios | Fox, Davis Entertainment, TSG Entertainment | Jolie Pas, Legendary Entertainment |
|---|---|---|---|
| Box Office ($ 100 million) | 8 | 1.6 | 1.6 |
| Movie series | yes | yes | no |
| How many movie series | 8 | 4 | 0 |
| Script | Drawn from Marvel Comics | Against alien invasion | A World War II story about survival, resilience and redemption |
| Experts' score | 20 | 3 | 0 |

After completing the annotation of the dataset, the movie dataset not only contains the basic data of movie samples but also gives the mentor signals of different movie samples, that is, the evaluation of the value of movie merchandising, which meets the requirements of supervised learning algorithms.

## 3.2. Data Analysis

### 3.2.1. Data distribution and relational analysis

After analyzing the data set samples, after the Delphi method analysis, the evaluation results of movie merchandising range from 0 to 25 points, and the data distribution is shown in Figure 3-3.:

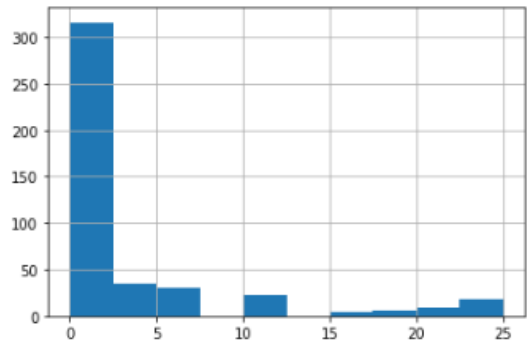

Figure 3-3 Film Merchandising Evaluation Results

The amount of sample data with movie merchandising value score of 0 is very large in all samples, which is a disadvantage. The relationship between movie ratings and the assessed value of movie merchandising is as Figure 3-4:

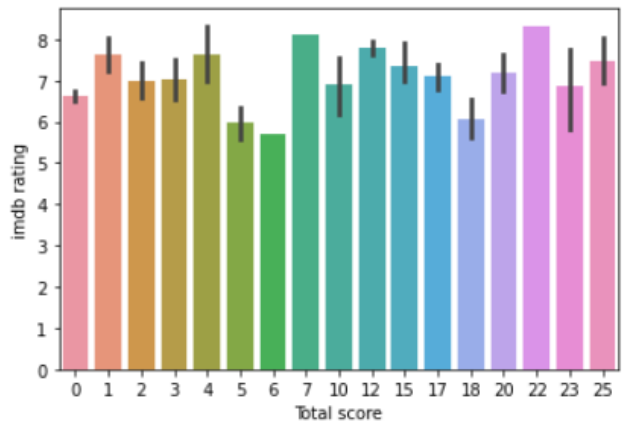

Figure 3-4 Film Scoring and Evaluation Value of Film Merchandising

The movie samples we select generally have global influence, so the ratings are basically high, but we can see from the above figure that movie ratings have little to do with the value of movie merchandising.

The relationship between the movie box office and the appraised value of movie merchandising is as Figure 3-5:

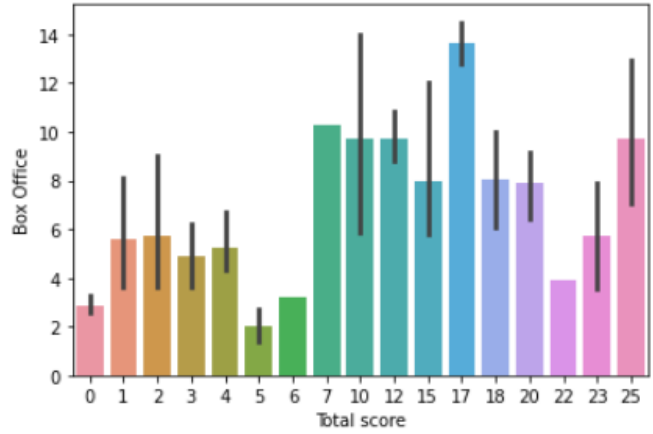

Figure 3-5 Movie Box Office and the Appraisal Value of Movie Merchandising

The relationship between whether a movie is a series and the number of series and the appraised value of movie merchandising is as Figure 3-6 and Figure 3-7.:

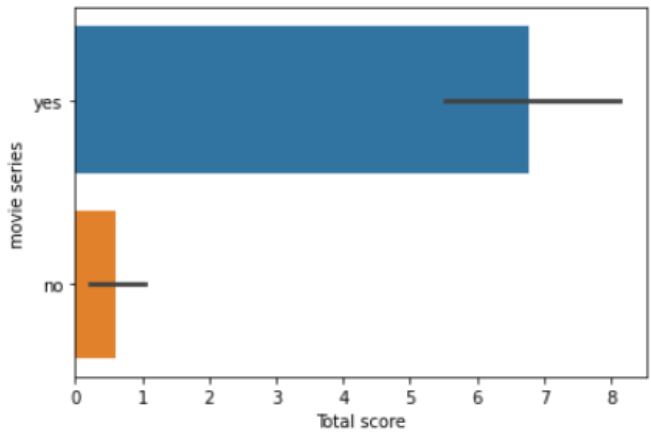

Figure 3-6 Whether the film is a series and film Merchandising evaluation value

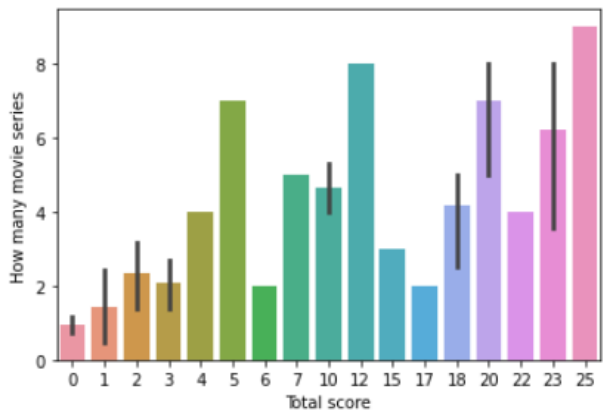

Figure 3-7 The number of series films and the evaluation value of film Merchandising

It can be seen from this that the merchandising value of series films far exceeds that of non-series films, and the merchandising values of films with a large number of series are generally higher. From the above analysis, we can roughly analyze the importance of various data for the evaluation of movie merchandising.

### 3.2.2. Handling of missing data

It often happens that some data in the dataset is missing. If the missing fields are importantly related to the evaluation of movie merchandising, it will inevitably affect the training effect. Therefore, we perform a null check on the dataset and the result is shown in Table 3-5

```
 #   Column                      Non-Null Count   Dtype
---  ------                      --------------   -----
 0   id                          441 non-null     int64
 1   film                        441 non-null     object
 2   year                        441 non-null     int64
 3   Motion Picture Rating(MPAA) 441 non-null     object
 4   time                        441 non-null     object
 5   imdb rating                 441 non-null     float64
 6   Grnres                      441 non-null     object
 7   Directors                   441 non-null     object
 8   writers                     421 non-null     object
 9   stars                       441 non-null     object
 10  countries of origina        441 non-null     object
 11  language                    441 non-null     object
 12  Filming locations           431 non-null     object
 13  Production Companies        441 non-null     object
 14  Box Office                  430 non-null     float64
 15  movie series                437 non-null     object
 16  How many movie series       437 non-null     int64
 17  Script                      441 non-null     object
 18  Total score                 441 non-null     int64
```

Table 3-5 Null check result

After inspection, it was found that some data in the dataset were missing, of which 10 were missing in Filming locations; in particular, the fields that were important in the evaluation of movie merchandising, such as Box Office, had only 430 and 11 were missing; movie series and How many movie series were missing 4 All missing data were filled in by contacting industry experts.

## 3.3. Data segmentation

### 3.3.1. Sample segmentation

In machine learning, generally speaking, we can't use all the data to train a model, otherwise, we will not have the data set to validate the model and evaluate the prediction effect of our model.

Divide the movie sample data set into a training set and a test set. As mentioned above, the number of evaluation results in the movie sample data is large. For example, the random segmentation method may cause too many or too few samples in the test set that are evaluated as 0, which affects the verification effect. Therefore, when dividing the data, we randomly select a certain amount of data from each score file to form a test set. The number of samples in the test set accounts for about 20% of the total number of samples, and the rest is used as the training set. In this way, the test set contains data of various evaluation grades, and the distribution of scores in the test set is basically consistent with the training set, so as to test the model training results.

### 3.3.2. Cross-validation analysis

Since the movie sample data is small, cross-training is used for validation during model training. We use the KFold function in the sklearn package to perform k-Fold Cross-Validation cross-validation. The dataset a is randomly divided into k packages, one of which is used as the test set each time, and the remaining k-1 packages are used as the training set. to train. Cross-validation dataset partitioning method is shown in Figure 3-8.

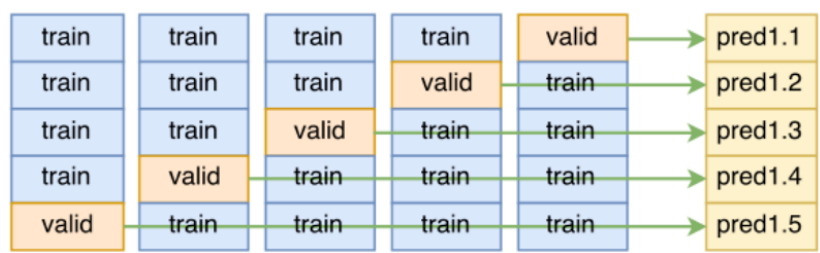

Figure 3-8 Cross-validation analysis

Divide the original data into K parts, K should not be too large or too small, as shown in Figure 3-9, we divide the data into 5 parts in the experiment. Perform K experiments in sequence, and each time one of the data is used as the validation set, the remaining K-1 are all used as the training set to train the model, and K evaluation values of the algorithm are obtained. The K evaluations are averaged to obtain a more objective performance evaluation of the model algorithm. Finally, all K pieces of data are used for training to obtain the final model. In this way, all data is used to train the model and an objective evaluation of the algorithm is also obtained, making full use of limited data in the case of a small amount of data.

## 3.4. Model training

As shown in Figure 4.1 Experimental flowchart, in the model training, we first used LightGBM, XGBoost, and LASSO three algorithm models to train at the same time, and then combined and integrated the WeightedEnsemble method to improve the prediction accuracy. One of the movie samples information example is shown in Table 3-6.

Table 3-6 Movie sample information example

| Movie | Year | Motion Picture Rating(MPAA) | Time | IMDB rating | |
|---|---|---|---|---|---|
| *Predator 2* | 1990 | R | 1h48m | 6.3 | |
| | **Genres** | **Directors** | **writers** | **stars** | |
| | Action, Horror, Sci-Fi | Stephen Hopkins | Jim Thomas, John Thomas | Danny Glover, Gary Busey, Kevin Peter Hall | |
| | **Countries of origina** | **Languages** | **Filming locations** | **Production Companies** | |
| | United States | English | BART, San Francisco, California, USA | Davis Entertainment, Lawrence Gordon Productions, Silver Pictures | |
| | **Box Office ($ 100 million)** | **movie series** | **How many movie series** | **Script** | |
| | 0.57 | yes | 4 | characters | |
| | **Experts' score** | **XGBoost** | **LASSO** | **LightGBM** | **WeightedEnsemble** |
| | 3 | 2 | 2 | **3** | **3** |

Take the movie *Predator 2* as an example, the above table is the sample information of the movie, the sample is a series, and generally has a good market value of merchandising, but because the movie is an R-rated Horror type, the audience is limited, and experts have no opinion on the value of movie merchandising. The score is 3, the result predicted by the machine learning algorithm is that the XGBoost and LASSO algorithms predict incorrectly, and the LightGBM and WeightedEnsemble predictions are correct.

In the movie sample, there are many samples with the merchandising value of 0. Table 3-7 shows an example of a movie merchandising with a value score of 0.

Table 3-7 movie merchandising with a value score of 0

| movie | year | Motion Picture Rating(MPAA) | time | IMDB rating | |
|---|---|---|---|---|---|
| *Thirteen Days* | 2000 | PG-13 | 2h25m | 7.3 | |
| | **Genres** | **Directors** | **Writers** | **Stars** | |
| | Drama,History,Thriller | Roger Donaldson | David SelfErnest R. May | Kevin Costner,Bruce Greenwood,Ernest R. May | |
| | **countries of origina** | **Languages** | **Filming locations** | **Production Companies** | |
| | United States | English,Russian,Spanish,Romanian | United States | New Line Cinema,Beacon Communications,Beacon Pictures | |
| | **Box Office($100 million)** | **movie series** | **How many movie series** | **Script** | |
| | 0.6 | no | 0 | | |
| | **Experts' score** | **XGBoost** | **LASSO** | **LightGBM** | **WeightedEnsemble** |
| | 0 | 0 | 0 | 0 | 0 |

The above table is the movie sample information of the movie *Thirteen Days*. This sample is not a series, and the merchandising value is usually not high. The expert's score for the value of the movie merchandising is 0. The four machine learning algorithms predict all correct. It has been observed that since there are many samples with the merchandising value of 0 in the movie sample, the prediction accuracy of the merchandising value of such movies is relatively high.

Table 3-8 is a comparison of the merchandising prediction results of a few more movie samples.

Table 3-8 Comparison of the merchandising prediction results

| movie | Experts' score | XGBoost | LASSO | LightGBM | Weighted Ensemble |
|---|---|---|---|---|---|
| *The Predator* | 3 | 2 | 2 | 3 | 3 |
| *Prometheus* | 3 | 2 | 2 | 3 | 3 |
| *Alien covenant* | 3 | 2 | 2 | 3 | 3 |
| *Terminator3: Rise of the Machines* | 5 | 3 | 4 | 3 | 4 |
| *Terminator: Dark Fate* | 5 | 1 | 0 | 0 | 1 |
| *Dark Phoenix* | 6 | 2 | 0 | 0 | 1 |
| *Deadpool* | 2 | 1 | 2 | 2 | 2 |
| *Thirteen Days* | 0 | 0 | 0 | 0 | 0 |
| *Undisputed* | 1 | 1 | 0 | 1 | 1 |
| *S.W.A.T.* | 0 | 0 | 0 | 0 | 0 |
| *The Texas Chainsaw Massacre* | 1 | 1 | 0 | 1 | 1 |
| *Stuck on You* | 0 | 0 | 0 | 0 | 0 |

It can be seen from the comparison of the above table that WeightedEnsemble is obtained by the weighted integration of three machine learning algorithms (described in detail below), so the accuracy rate is higher than that of independently used machine learning algorithms, which is also one of the important achievements of this paper.

# 4. Model Results Analysis

## 4.1. Algorithm Verification Analysis

After various model algorithms are trained, we use the test set data for verification. The test set data does not contain expert evaluation annotations. The algorithm model obtained by training predicts the value of movie merchandising based on the test set data, and then the predicted data $pred_i$ and Experts evaluate data scores for comparison if:

$$|\text{pred}_i - score_i| <= 1\% * score_i$$

Then it is considered that the algorithm model is accurate in predicting the `merchandising` value of the i-th movie in the test set. The four algorithm model prediction verification statistical results obtained from this are as Table 4-1:

Table 4-1 Four Algorithms Model Prediction Verification Results

| algorithm | Linear | LightGBM | LASSO | XGBoost |
|---|---|---|---|---|
| **Accuracy** | 43.75% | **62.50%** | 58.75% | 57.50% |

It can be seen from Table 4-1 that the Linear model has the lowest prediction accuracy, only 43.75%, and the LightGBM model has the highest prediction accuracy, reaching 62.50%. The LightGBM model is optimized from the XGBoost model, and it performs better in the case of small sample data. it is good.

## 4.2. WeightedEnsemble model

Through the analysis of the algorithm verification results, we found that the prediction accuracy of Linear algorithm is only 43.75%, which can be discarded; while the prediction accuracy of LightGBM, LASSO, XGBoost three algorithm models are relatively close, the maximum is 62.50%, the minimum is 57.50%, the difference is only 5%, after careful observation of the prediction results, it is found that the correct movie samples predicted by these three algorithm models are different. For example, for the movie *101 Dalmatians*, the LASSO algorithm model predicts accurately, while the LightGBM and XGBoost algorithms predict incorrectly. There is an idea that the prediction results of the three algorithm models are integrated into a new algorithm model, which has proved the effect in many experiments [23][24]. In this paper, the weighted integration method is adopted, so we name the weighted integration model WeightedEnsemble, or WE model for short, and the integration method is as follows:

$$WE\text{-}pred = \alpha_1 * LightGBM\text{-}pred + \alpha_2 * LASSO\text{-}pred + \alpha_3 * XGBoost\text{-}pred$$

Among them, LightGBM-pred, LASSO-pred, and XGBoost-pred are the predicted values of the three algorithm models of LightGBM, XGBoost, and LASSO, respectively; $\alpha_1, \alpha_2, \alpha_3$ are the weighting coefficients $\alpha_1, \alpha_2, \alpha_3$ The value is between 0-1, And $\alpha_1+\alpha_2+\alpha_3 = 1$.

The value of the weighting coefficient, the initial value is given randomly and is continuously adjusted and optimized according to the obtained prediction and verification accuracy, and finally, the optimal combination is obtained. The accuracy of the prediction and verification is as Table 4-2:

Table 4-2 Accuracy of the prediction and verification

| algorithm | Linear | LightGBM | LASSO | XGBoost | WeightedEnsemble |
|---|---|---|---|---|---|
| Accuracy | 43.75% | 62.50% | 58.75% | 57.50% | **72.50%** |

It can be seen that the prediction accuracy after the optimization of the weighted ensemble model algorithm reaches 72.50%, which is 10% higher than the accuracy of the original best performing LightGBM model.

## 4.3. Conclusion

In this paper, the innovation proposes to study movie merchandising in a "quantitative" way, which can make up for the shortcomings of qualitative research and promote the development of the film industry. This research pioneered a machine learning algorithm model——WE model for movie merchandising market forecasting. Through an extensive collection of movie data, a set of datasets for movie merchandising research was created, and a large number of industry experts were invited to annotate movie datasets. The Delphi method is used to evaluate and score the movie merchandising of the sample data, so that the movie data set has real mathematical research value. Finally, the WE model is used to train and predict the algorithm model of the labelled movie data set, and the prediction accuracy rate reaches 72.50%. It is proved that the machine learning algorithm is effective in the prediction of movie merchandising. It can predict the value of the surrounding products before the movie is released and create greater profits for the movie products, which will play a major role in the development of the movie industry. Similarly, the algorithm proposed in this paper can also be applied to the merchandising value prediction of animation movies and animated TV series. In the follow-up research, we will continue to deepen the application scope of the WE model and improve the prediction accuracy.

# Reference

[1] Ravid, S. A. (2005). The Big Picture: The New Logic Of Money And Power In Hollywood,

[2] Justin Wyatt, High Concept: Movies and Marketing in Hollywood, (Austin: University of Texas Press, 1994), 148.

[3] Samantha Loveday (2020). Global sales of licensed goods up 4.5% to US$292.8 billion.https://www.licensingsource.net/global-sales-of-licensed-goods-up-4-5-to-us292-8-billion/

[4]License Global. https://www.licenseglobal.com/rankings-and-lists/top-150-leading-licensors.2021-06-25.

[5]Gaines, J. (1989). The queen Christina tie‐ups: Convergence of show window and screen. Quarterly Review of Film & Video, 11(1), 35-60.

[6]Eckert, C. (1978). The Carole Lombard in Macy's Window. Quarterly Review of Film & Video, 3(1), 1-21.

[7]Hills, M. (2003). Fan cultures. London: Routledge.p.27.

[8]Jonathan Gray（2010). Show Sold Separately: Promos, Spoilers, and Other Media

[9]Johnson, D. (2014). "May the Force be with Katie" Pink media franchising and the postfeminist politics of HerUniverse. Feminist media studies, 14(6), 895-911.

[10] Horváth, Á., & Gyenge, B. (2018). Movie merchandising and its consumer perception. Management, 16, 18.

[11] Corazza, A., Maggio, V. & Scanniello, G. Coherence of comments and method implementations: a dataset and an empirical investigation. Software Qual J 26, 751–777 (2018).

[12] Kureljusic, M., Karger, E. Data Preprocessing as a Service – Outsourcing der Datenvorverarbeitung für KI-Modelle mithilfe einer digitalen Plattform. Informatik Spektrum 45, 13–19 (2022).

[13] Sarker, I.H. Machine Learning: Algorithms, Real-World Applications and Research Directions. SN COMPUT. SCI. 2, 160 (2021).

[14] Sharma, D., Kumar, B., Chand, S. et al. A Trend Analysis of Significant Topics Over Time in Machine Learning Research. SN COMPUT. SCI. 2, 469 (2021).

[15] Zhang, D., Tsai, J.J. Machine Learning and Software Engineering. Software Quality Journal 11, 87–119 (2003).

[16] Pandit, P., Dey, P. & Krishnamurthy, K.N. Comparative Assessment of Multiple Linear Regression and Fuzzy Linear Regression Models. SN COMPUT. SCI. 2, 76 (2021).

[17] Rajan, M.P. An Efficient Ridge Regression Algorithm with Parameter Estimation for Data Analysis in Machine Learning. SN COMPUT. SCI. 3, 171 (2022).

[18] Ahmad, I.S., Bakar, A.A., Yaakub, M.R. et al. A Survey on Machine Learning Techniques in Movie Revenue Prediction. SN COMPUT. SCI. 1, 235 (2020).

[19] Lee K, Park J, Kim I, Choi Y. Predicting movie success with machine learning techniques: ways to improve accuracy. Inf Syst Front. 2018;20(3):577–88.

[20] Parimi R, Caragea D. Pre-release box-office success prediction for motion pictures. In: Lecture notes in computer science (including subseries lecture notes in artificial intelligence and lecture notes in bioinformatics). vol. 7988. LNAI; 2013. p. 571–85.

[21] Quader N, Gani MO, Chaki D, Ali MH. A machine learning approach to predict movie box-office success. In: 20th international conference of computer and information technology, ICCIT 2017, vols. 2018-Janua. Institute of Electrical and Electronics Engineers Inc.; 2018. p. 1–7

[22] Ruhrländer RP, Boissier M, Uflacker M. Improving box office result predictions for movies using consumer-centric models. In: KDD '18 proceedings of the 24th ACM SIGKDD international conference on knowledge discovery and data mining, KDD '18. London: ACM New York, NY, USA; 2018. p. 655–64.

[23] Cruz, J., Mamani, W., Romero, C. et al. Selection of Characteristics by Hybrid Method: RFE, Ridge, Lasso, and Bayesian for the Power Forecast for a Photovoltaic System. SN COMPUT. SCI. 2, 202 (2021).

[24] Rahul Automatic identification and classification of power quality events using a hybrid intelligent approach. Iran J Comput Sci 4, 115–124 (2021).